\documentclass[superscriptaddress,aps,prl,twocolumn,showpacs,preprintnumbers,amsmath,amssymb,floatfix,groupedaddress]{revtex4}

\usepackage{graphicx}
\usepackage{dcolumn}
\usepackage{bm}
\usepackage{color}

\begin{document}

\title{Photonic Phase Gate via an Exchange of 
Fermionic Spin Waves in a Spin Chain}

\author{Alexey V. Gorshkov}
\affiliation{Physics Department, Harvard University, Cambridge, Massachusetts 02138, USA}
\author{Johannes Otterbach}
\affiliation{Department of Physics and Research Center OPTIMAS, Technische Universit\"at Kaiserslautern, 67663, Kaiserslautern, Germany}
\author{Eugene Demler}
\affiliation{Physics Department, Harvard University, Cambridge, Massachusetts 02138, USA}
\author{Michael Fleischhauer}
\affiliation{Department of Physics and Research Center OPTIMAS, Technische Universit\"at Kaiserslautern, 67663, Kaiserslautern, Germany}
\author{Mikhail D. Lukin}
\affiliation{Physics Department, Harvard University, Cambridge, Massachusetts 02138, USA}

\date{\today}

\begin{abstract}

We propose a new protocol for implementing the two-qubit photonic phase gate. In our approach, the $\pi$ phase is acquired by mapping two single photons into atomic excitations with fermionic character and exchanging 
their positions.  
The fermionic excitations are realized as spin waves in a spin chain, while photon storage techniques provide the interface between the photons and the spin waves. Possible imperfections and experimental systems suitable for implementing the gate are discussed. 

\end{abstract} 

\pacs{03.67.Lx, 42.65.Hw, 03.67.Ac, 37.10.Jk}

\maketitle

Strong nonlinear interactions between single photons are essential for many potential applications in quantum information processing, such as efficient optical quantum computing and fast long-distance quantum communication  \cite{bouwmeester00,franson04}. 
However, since single-photon nonlinearities are generally very weak \cite{boyd07d}, strong photon-photon interactions require elaborate and experimentally challenging 
schemes    
 \cite{deutsch92,schmidt96,masalas04,muschik08}
with a reliable, practical approach yet to emerge.
In principle, a robust and conceptually
simple gate between two photons can be achieved  
by temporarily storing their quantum information into two excitations with fermionic character and exchanging these excitations to obtain the $\pi$-phase shift \cite{franson04}.
In this Letter, we propose how to implement this scheme with the role of fermionic excitations played by spin waves in a one-dimensional (1D) spin chain.

\begin{figure}[t]
\begin{center}
\includegraphics[width = \columnwidth]{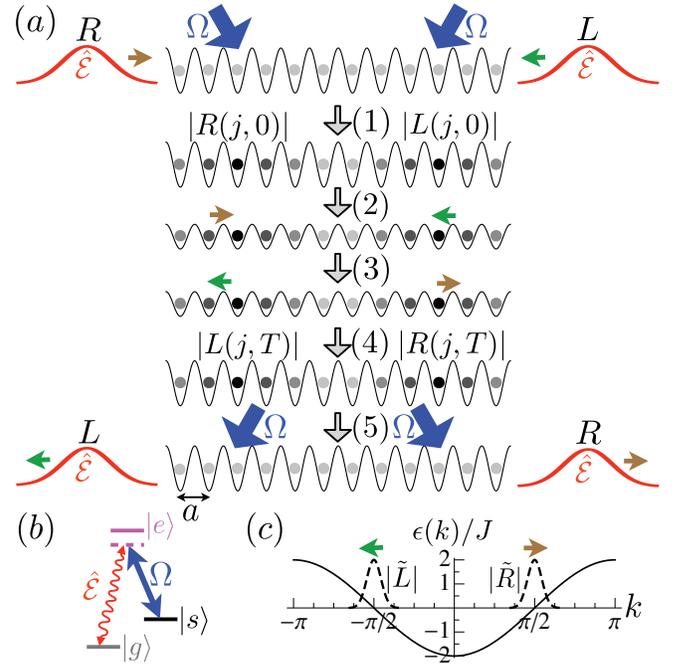}
\caption{(color online) (a) Schematic diagram for the implementation of the photonic two-qubit phase gate. 
(b) The $\Lambda$-type atomic level diagram used for interfacing photons and spin waves. (c) The fermionic dispersion relation $\epsilon(k)$ in units of $J$ (solid line), and, schematically, the amplitudes of the two spin waves in $k$-space (dashed lines).
\label{fig1}}
\end{center}
\end{figure}

The main idea of our protocol is illustrated schematically in Fig.~\ref{fig1}(a). Two photonic wave packets ($\hat{\mathcal{E}}$ in the Figure), 
labeled as $R$ (right-moving) and $L$ (left-moving), 
propagate in an optical  waveguide \cite{christensen08,bajcsy09,vetsch09} coupled to a 1D optical lattice filled with one bosonic $\Lambda$-type atom [Fig.~\ref{fig1}(b)] 
per site and with tunneling between sites turned off \cite{duan03}. In step (1), photon storage techniques \cite{fleischhauer02,lukin03,dutton04,gorshkov07} are used to map the two photons 
via an auxiliary control field (labeled $\Omega$) 
onto two spin waves formed by the collective coherence between atomic states $g$ and $s$ 
(the amplitude of the spin wave on a given atom is indicated by the darkness of the circle). In step (2), the lattice depth is reduced yielding a nearest-neighbor superexchange spin Hamiltonian \cite{duan03,trotzky08}, which couples 
states $s$ and $g$  and which can be adjusted in such a way that the spin waves behave as free fermions \cite{giamarchi03}. In step (3), the spin waves propagate through each other, exchange places, and, being fermionic, pick up the desired $\pi$ phase. In step (4), the superexchange is turned off. Finally, in step (5), the spin waves are retrieved back onto photonic modes.

The gate is thus achieved via an exchange of two free fermionic excitations that temporarily carry the photonic qubits. A closely related idea of endowing photons with fermionic character by means of  nonlinear 
atom-photon interactions is 
presented in Ref.~\cite{franson04}. 
Contrary to this, in our proposal, fermionic behavior of photons is achieved by mapping photonic qubits onto atomic states and then relying on 
atom-atom interactions. Such an approach is similar to that of Refs.\ \cite{masalas04,dutton04,muschik08}, 
where photonic quantum information is also processed by mapping it onto atomic states and using atom-atom interactions.

Since our gate is based on the relative positions of the fermions, it is robust in that the nonlinear phase is exactly $\pi$ (other errors are still possible, as discussed below). 
This attractive feature of our gate relies on the capability of using the presence and absence of a fermion as a qubit \cite{bravyi02, clark05}, which is in turn enabled by the interconversion between bosonic photons and fermionic spin waves \cite{franson04}.
This interconversion is achieved in two steps.
First, photon storage and retrieval 
techniques \cite{fleischhauer02,lukin03,dutton04,gorshkov07} are used to interconvert between photons and bosonic spin waves, and, second, 
hard-core interactions between spin waves effectively endow them with fermionic character.
The first step requires strong collectively enhanced coupling between photons and atoms \cite{fleischhauer02,lukin03,dutton04,gorshkov07}, which can be achieved in recently demonstrated waveguides that transversely confine both 
atoms and photons to dimensions on the order of a wavelength \cite{christensen08,bajcsy09,vetsch09}. 
The second step is achieved by means of strong atom-atom interactions \cite{duan03,trotzky08}. 
Our proposal thus bears a similarity to those of   
Refs.~\cite{masalas04,muschik08}, 
which also rely on strong interatomic interactions, 
but contrasts with methods where atom-photon interactions alone suffice  \cite{schmidt96,franson04}. 
At the same time, our proposal is fundamentally different from Refs.~\cite{masalas04,muschik08} 
in that it relies on the robustness and conceptual simplicity of fermionic exchange for achieving the nonlinear 
phase. 

Before proceeding, we note that fermionic exchange in spin chains, including their cold atom realization \cite{clark05},  
has 
been suggested   for use as an entangling gate in Refs.~\cite{clark05,yung05}. 
The advantage of our approach is that we require neither spatially inhomogeneous coupling nor single-site addressiblity 
and that we extend the gate from spins to photons via photon storage 
techniques \cite{fleischhauer02,lukin03,dutton04,gorshkov07}.

\textit{Details of the protocol.}---We begin with a bosonic two-component single-band Hubbard model in a 1D lattice, with the transverse motion frozen. 
There are $N$ sites with nearest-neighbor spacing $a$, yielding a total length $N a$. 
The tunneling amplitudes are $t_\alpha$ for species $\alpha$ ($\alpha = g$,$s$), while the s-wave interaction energies are $U_{\alpha \alpha}$ for two $\alpha$ atoms and $U_{sg}$ for an $s$ atom with a $g$ atom \cite{duan03}.
Assuming one atom per each of $N$ sites in the weak tunneling limit $t_g, t_s \ll U_{gg}, U_{ss}, U_{sg}$, ignoring edge effects, and dropping the term $\sum_j S^z_j$ (which would contribute a simple linear phase  \cite{phase}), the superexchange XXZ Hamiltonian is ($\hbar = 1$ throughout the Letter) \cite{duan03,trotzky08}
\begin{eqnarray}
H = \sum_{j = 1}^{N-1} \left[ - J (S_j^+ S_{j+1}^- + S_j^- S_{j+1}^+)+ V S_j^z S_{j+1}^z\right], \label{H0}
\end{eqnarray}
where $V = 2 \frac{t_g^2 + t_s^2}{U_{sg}} - \frac{4 t_g^2}{U_{gg}} - \frac{4 t_s^2}{U_{ss}}$ and $J = \frac{2 t_g t_s}{U_{sg}}$.
Here $\vec S_j = \vec \sigma_j/2$, where $\vec \sigma_j$ are Pauli operators in the $\{s,g\}$ 
basis and $S^\pm_j = S^x_j \pm i S^y_j$. It is convenient to use the Jordan-Wigner transformation \cite{giamarchi03} to map the spins onto fermions with creation operators $c^\dagger_j$: $S^+_j = c^\dagger_j \exp\left(i \pi \sum_{k = 1}^{j-1} n_k\right)$ and $S^z_j = n_j - \frac{1}{2}$, where $n_j = c^\dagger_j c_j$. Then, dropping the term $\sum_j n_j$, which would contribute only a linear phase \cite{phase},
\begin{eqnarray}
H = - J \sum_j \left(c^\dagger_j c_{j+1} + c^\dagger_{j+1} c_j\right)+ V \sum_j n_j n_{j+1}. \label{H1}
\end{eqnarray}
With $V = 0$, Eq.~(\ref{H0}) is the XX Hamiltonian, while Eq.~(\ref{H1}) then represents free fermions. We can obtain $V = 0$, for example, by taking $t_g = t_s = t$ and $U_{gg} = U_{ss} = 2 U_{sg} = U$, so that $J  = 4 t^2/U$ \cite{clark05}. When all three scattering lengths are similar in size, as e.g.~in some ground-state alkali atoms, $U_{sg}$ can be reduced by shifting the $s$ and $g$ lattices relative to each other \cite{mandel03}.

The free fermion ($V = 0$) case can be diagonalized, ignoring boundaries, by going to $k$-space:  
\begin{equation}
H = \sum_k \epsilon(k) c^\dagger_k c_k, \label{Hk}
\end{equation}
where $k$ takes $N$ values in the first Brillouin zone from $-\pi$ to $\pi$ at intervals of $2 \pi/N$, 
$c^\dagger_k = (1/\sqrt{N}) \sum_j e^{i k j} c^\dagger_j$,  and the 
dispersion $\epsilon(k) = - 2 J \cos(k)$ is shown in Fig.~\ref{fig1}(c). 

With all atoms in $|g\rangle$ (denoted as $|\textrm{vac}\rangle$) and tunneling turned off, two counterpropagating photonic modes, labeled as $R$ and $L$ in Fig.~\ref{fig1}(a), are incident on our 1D chain of atoms near resonance with the $g$-$e$ transition [see Fig.~\ref{fig1}(b)].  
The wave packets carry photonic qubits,  in which $|0\rangle$ ($|1\rangle$) corresponds to the absence (presence) of a single photon in the mode. The goal is to implement a two-qubit phase gate, which puts a $\pi$ phase on the state $|1\rangle |1\rangle$, leaving the other three basis states ($|0\rangle|0\rangle$, $|0\rangle|1\rangle$, and $|1\rangle|0\rangle$) unchanged \cite{phase}. 
We use two control fields $\Omega$ (one for $R$ and one for $L$) near resonance with the $s$-$e$ transition [see Fig.~\ref{fig1}(b)] applied at different angles from the side of the chain  [see Fig.~\ref{fig1}(a)]  
to store  \cite{fleischhauer02,lukin03,dutton04,gorshkov07} the two photons into the $s$-$g$ coherence over two spatially separated regions. 
As shown below, 
by appropriately choosing the angle of incidence of the control fields, we can arrange for the 
$R$ and $L$ 
spin waves to be centered in $k$-space around $k = \pi/2$ and $k = - \pi/2$, respectively, as shown in Fig.~\ref{fig1}(c). This ensures that the two spin waves will move through each other with maximum speed [largest possible difference of first derivatives of $\epsilon(k)$]  and with minimal distortion [vanishing second derivative of $\epsilon(k)$]. Photonic band gap effects coming from the periodicity of atomic positions \cite{petrosyan07} can be avoided if light of different wavelengths is used for the lattice and the photons, which is typically the case in experiments \cite{schnorrberger09}. %

The XX Hamiltonian is then turned on. Since $H$ is diagonal in the Fourier basis [Eq.~(\ref{Hk})], it is convenient to define, for any function $q(j)$ on the sites $j$, its Fourier transform $\tilde q(k) = (1/\sqrt{N}) \sum_j q(j) e^{-i k j}$ together with the usual inverse relation $q(j) = (1/\sqrt{N}) \sum_k \tilde q(k) e^{i k j}$. Using Eq.~(\ref{Hk}) and $S^+_j |\textrm{vac}\rangle = c^\dagger_j |\textrm{vac}\rangle$, it is then easy to check that the $R$ spin wave alone evolves in time $\tau$ as $\sum_{j} R(j,\tau) S_+^j |\textrm{vac}\rangle$, where $\tilde R(k,\tau) = \tilde R(k,0) e^{- i \tau \epsilon(k)}$. We now assume that $R(j,0)$ is centered around $j = N/4$ and has width $\sim N$ that is nevertheless small enough for $R(j,0)$ to be negligible near the edge ($j = 1$) and in the middle ($j = N/2$) of the chain. Furthermore, we assume that $\tilde R(k,0)$ is centered around $k = \pi/2$ with Fourier-transform-limited width $\sim 1/N$, so that, for $N \gg 1$, the dispersion is approximately linear for all relevant $k$: $\epsilon(k) \approx [k - (\pi/2)] v$, where the velocity (in sites per second) is $v = 2 J$. Then, provided the pulse does not reach $j = N$, it moves to the right with velocity $v$: $R(j,\tau) \approx e^{i \tau J \pi} R(j - v  \tau,0)$.  Such spin-wave propagation plays an important role in quantum information transport in spin chains \cite{osborne04}. Similarly, the spin wave $L(j,0)$ centered around $j = 3 N/4$ with carrier momentum $- \pi/2$ approximately evolves as $L(j',\tau) \approx e^{i \tau J \pi} L(j' + v \tau,0)$. Finally, if both $R$ and $L$ are stored simultaneously, they propagate as $\sum_{j} R(j,\tau) c^\dagger_{j} \sum_{j'} L(j',\tau) c^\dagger_{j'} |\textrm{vac}\rangle$, so that after time $\tau = T \equiv N/(2 v)$ they exchange places. However, in addition, when rewriting the final state in terms of spin operators $S^+_j$, an extra minus sign appears since for all relevant $j$ and $j'$, $j > j'$. Thus, the state $|1\rangle |1\rangle$ picks up an extra $\pi$ phase relative to the other three basis states, giving rise to the two-qubit photonic phase gate once the spin waves are retrieved. 

An alternative way to see the emergence of the minus sign is to note that, for $V = 0$ and ignoring boundaries, the eigenstates of the two-excitation sector of Eq.~(\ref{H0}) are $\propto \sum_{j < j'} \left(e^{i k j} e^{i p j'} - e^{i p j} e^{i k j'}\right) |j,j'\rangle$ 
and have energy $\epsilon(k) + \epsilon(p)$, where $p$ and $k$ are quantized as before, $p < k$, and $|j,j'\rangle = S^+_{j} S^+_{j'} |\textrm{vac}\rangle$. The simultaneous propagation of $R$ and $L$ spin waves then explicitly exhibits the minus sign: $\sum_{j < j'} \left[R(j,\tau) L(j',\tau) - R(j',\tau) L(j,\tau)\right] |j,j'\rangle$,
so that at $\tau = 0$ ($\tau = T$) only the first (second) term in the square brackets contributes.

\textit{Experimental realizations.}---Two experimental systems well suited for the implementation of our phase gate are atoms confined in a hollow-core photonic band gap fiber \cite{christensen08,bajcsy09} and atoms trapped in the evanescent field around an ultrathin optical fiber \cite{vetsch09}. In the former system, a running-wave red-detuned laser can be used to provide a 
transverse potential limiting atomic motion to a tube and preventing collisions with fiber walls. 
Then either a blue-detuned \cite{schnorrberger09} or another red-detuned beam can be used to create a 1D lattice in the tube. To prepare the atoms, one can load a Bose-Einstein condensate into the fiber with the lattice turned off (but the tube confinement on), and then adiabatically turn on the lattice bringing the atoms via a phase transition into the Mott state \cite{weld09}.
Recent experiments indicate a temperature of $U/(37 k_B)$ \cite{weld09}, at which, with a properly adjusted density, a state of one atom per each of $N = 1000$ sites would be defect-free \cite{gerbier07}. In fact, $> 99\%$ probability of single-site unit occupancy has already been demonstrated in a Mott insulator \cite{bakr10}.
Before loading, optical pumping and state selective trapping can be used to prepare all the atoms in state $g$. 
It is important to note that, since the initial spin state is determined by optical pumping, the protocol does not require 
the temperature of the original atomic cloud to be below $J/k_B$ \cite{trotzky08}.
The same procedure can be used to load the atoms in the evanescent field system.

In both experimental systems, during photon storage, an incoming photon of momentum $\vec k_i$ (parallel to the atomic chain axis) 
is absorbed while a control field photon of momentum $\vec k_c$ is emitted. The $k$-vector of the spin wave is, thus, equal to the projection on the atomic chain axis of $\vec k_i - \vec k_c$  \cite{fleischhauer02}.
For example, 
if $|\vec k_i| \approx |\vec k_p| \approx \pi/a$, then an angle \cite{angle} 
of $\approx 60^\circ$ between the control beam and the atomic chain axis gives the desired spin-wave $k$-vector of $\pm \pi/(2 a)$. 
For $N = 1000$ and $2 a \sim 1 \mu$m, the length of the medium is $N a \sim 500 \mu$m.  
Thus, one could indeed use two focused control beams to store independently two single photons from the opposite directions. One could also store photons incident from the same direction, in which case additional Raman transitions or gradients in Zeeman or Stark shifts may be used to produce the desired spin-wave momenta. Spin-wave retrieval is carried out in the same manner. 

\textit{Imperfections.}---We now consider several errors that can arise during gate execution. First, to estimate the error due to the finite $t/U$ ratio, we perform two consecutive Shrieffer-Wolff transformations to compute the $t^4/U^3$ corrections to $H$. 
A perturbative calculation then shows that the dominant effect of these corrections, beyond a slight and unimportant modification of the dispersion $\epsilon(k)$, is an additional nonlinear phase $\sim (t/U)^2$. This yields an error $p_1 \sim (t/U)^4$, which can be further reduced by tuning $V$ (see Conclusion). 
Second, photon storage and retrieval with error $p_2 \sim 1/(\eta N)$ \cite{gorshkov07c} can be achieved  at any detuning \cite{gorshkov07} and for pulse bandwidths as large as $\sim \eta N \Gamma$ 
\cite{gorshkov08b}, where $\eta N$ is the resonant optical depth on the $e$-$g$ transition whose linewidth is $\Gamma$. 
Third, the error due to the decay of the $s$-$g$ coherence with rate $\gamma_0$ is $p_3 \sim \gamma_0 T$. 
An additional error comes from the reshaping of the pulse due to the nonlinearity of the dispersion. 
This error falls off very quickly with $N$, and already for $N = 100$ we find it to be as low as $\sim 10^{-4}$. 
Moreover, pulse shape distortion can be further corrected during retrieval \cite{gorshkov07,gorshkov07c,novikova08}, making the corresponding error negligible.

With an experimentally demonstrated $\eta = 0.01$ \cite{bajcsy09,vetsch09}, we need $N \gtrsim 1000$ to achieve efficient photon storage and retrieval (small $p_2$). 
To suppress $t^4/U^3$ corrections to $H$, we take $(t/U)^2 = 0.01$, which reduces $p_1$ down to $\lesssim 10^{-4}$ and yields velocity $v = 8 t^2/U$ and propagation time $T = N/(2 v) = N/(0.16 U)$. For $U = (2 \pi) 4$ kHz \cite{fukuhara09} and $N = 1000$, this gives $T \sim 250$ ms, which is shorter than the experimentally observed coherence times of $\sim 1$ s \cite{zhang09b,schnorrberger09}.
Thus, a proof-of-principle demonstration of our gate can 
be carried out with current experimental technology.
With improved experimental systems, a faster and higher fidelity implementation will 
be possible. In particular, coherence times and $\eta$ can likely be improved with better control of light and atoms. 
Moreover, larger $U$ might be obtainable 
via magnetic or optical Feshbach resonances \cite{tiesinga93ciurylo05} or with more intense lattice lasers. 

For an incoming single photon of duration $T_p = 100$ ns \cite{eisaman05}, using $\eta N = 10$ and the parameters of the ${}^{87}$Rb D$_1$ line, the required peak control Rabi frequency is $\Omega \sim \sqrt{\eta N \Gamma/T_p} \approx (2 \pi) 10$ MHz \cite{gorshkov07c}, corresponding to $<$ 100 $\mu$W of power for a beam diameter of 200 $\mu$m. For a frequency difference of several GHz between the quantum and the control fields, stray control light can be easily filtered out \cite{eisaman05, bajcsy09}.

\textit{Conclusion.}---We have proposed a robust photonic phase gate based on the exchange of two fermionic excitations that temporarily carry the photonic qubits. 
One of the advantages of our protocol is that, as in Ref.~\cite{muschik08}, the spin chain can be simultaneously used not only to couple but also to store the photonic qubits, which is crucial for many quantum information processing tasks \cite{zoller05}. While we have described how the gate works in the occupation basis, it can easily be extended to the more convenient polarization basis \cite{bouwmeester00} 
simply by applying the above gate to just one of the two polarizations. We also envision extensions to fermionic atoms (which will ease initial state preparation \cite{viverit04}), to Coulomb-coupled ions enclosed in a cavity \cite{herskind09b}, 
and to dipole-dipole interacting solid-state emitters (such as quantum dots or nitrogen-vacancy color centers in diamond) coupled to surface plasmons in conducting nanowires 
\cite{chang06}.
Finally, as will be shown elsewhere, the case $V \neq 0$ gives access to a phase gate with a tunable phase given by $\pi - 2 \tan^{-1}\left[V/(2 J)\right]$. With current 
experimental systems already sufficient for a proof-of-principle demonstration,  
our protocol should be immediately useful 
as a gate in quantum 
computation and communication 
and as a probe of spin chain physics.

We thank M.~Metlitski,  W.~Bakr, G.~Nikoghosyan, M.~Bajcsy, S.~Hofferberth, M.~Hafezi, P.~Rabl, J.~Thompson, M.~Gullans, N.~Yao, J.~Brask, S.~Dawkins, and E.~Vetsch for discussions. This work was supported by NSF, CUA, DARPA, AFOSR MURI, and the Deutsche Forschungsgemeinschaft (DFG) through the GRK 792.



\begin{thebibliography}{51}
\bibitem{angle} One can apply control light at an angle to the hollow-core fiber since the band gap is only partial \cite{knight98}.
\end{thebibliography}

\end{document}